\documentclass[aip,superscriptaddress,amsmath,amssymb,floatfix,reprint,twocolumn]{revtex4-1}
\pdfoutput=1

\usepackage{amssymb}
\usepackage{graphicx}
\usepackage{bm}
\usepackage[pdftex,breaklinks=true,bookmarksopen=true,bookmarksopenlevel=3,bookmarksnumbered=true,colorlinks=true,urlcolor= magenta,citecolor=blue,linkcolor=blue]{hyperref}
\usepackage{float}
\usepackage[utf8]{inputenc}
\usepackage[T1]{fontenc}
\usepackage{verbatim}
\usepackage{siunitx}
\usepackage[dvipsnames]{xcolor}
\usepackage{soul}
\usepackage{ulem}

\begin{document}
\title{Probing strong-field QED in beam-plasma collisions}

\author{A. Matheron}
\thanks{These authors have contributed equally to this work.}
\affiliation{LOA, ENSTA Paris, CNRS, Ecole Polytechnique, Institut Polytechnique de Paris, 91762 Palaiseau, France}
\author{P. San Miguel Claveria}
\thanks{These authors have contributed equally to this work.}
\affiliation{LOA, ENSTA Paris, CNRS, Ecole Polytechnique, Institut Polytechnique de Paris, 91762 Palaiseau, France}
\author{R. Ariniello}
\affiliation{University of Colorado Boulder, Department of Physics, Center for Integrated Plasma Studies, Boulder, Colorado 80309, USA}
\author{H. Ekerfelt}
\affiliation{SLAC National Accelerator Laboratory, Menlo Park, CA 94025, USA}
\author{F. Fiuza}
\affiliation{SLAC National Accelerator Laboratory, Menlo Park, CA 94025, USA}
\author{S. Gessner}
\affiliation{SLAC National Accelerator Laboratory, Menlo Park, CA 94025, USA}
\author{M F. Gilljohann}
\affiliation{LOA, ENSTA Paris, CNRS, Ecole Polytechnique, Institut Polytechnique de Paris, 91762 Palaiseau, France}
\author{M. J. Hogan}
\affiliation{SLAC National Accelerator Laboratory, Menlo Park, CA 94025, USA}
\author{C.~H.~Keitel}
\affiliation{Max-Planck-Institut f\"ur Kernphysik, Saupfercheckweg 1, D-69117 Heidelberg, Germany}
\author{A. Knetsch}
\affiliation{LOA, ENSTA Paris, CNRS, Ecole Polytechnique, Institut Polytechnique de Paris, 91762 Palaiseau, France}
\author{M.~Litos}
\affiliation{University of Colorado Boulder, Department of Physics, Center for Integrated Plasma Studies, Boulder, Colorado 80309, USA}
\author{Y.~Mankovska}
\affiliation{LOA, ENSTA Paris, CNRS, Ecole Polytechnique, Institut Polytechnique de Paris, 91762 Palaiseau, France}
\author{S.~Montefiori}
\affiliation{Max-Planck-Institut f\"ur Kernphysik, Saupfercheckweg 1, D-69117 Heidelberg, Germany}
\author{Z. Nie}
\affiliation{University of California Los Angeles, Los Angeles, CA 90095, USA}
\author{B. O'Shea}
\affiliation{SLAC National Accelerator Laboratory, Menlo Park, CA 94025, USA}
\author{J.~R.~Peterson}
\affiliation{SLAC National Accelerator Laboratory, Menlo Park, CA 94025, USA}
\affiliation{Stanford University, Physics Department, Stanford, CA 94305, USA}
\author{D. Storey}
\affiliation{SLAC National Accelerator Laboratory, Menlo Park, CA 94025, USA}
\author{Y.~Wu}
\affiliation{University of California Los Angeles, Los Angeles, CA 90095, USA}
\author{X. Xu}
\affiliation{SLAC National Accelerator Laboratory, Menlo Park, CA 94025, USA}
\author{V. Zakharova}
\affiliation{LOA, ENSTA Paris, CNRS, Ecole Polytechnique, Institut Polytechnique de Paris, 91762 Palaiseau, France}
\author{X. Davoine}
\affiliation{CEA, DAM, DIF, 91297 Arpajon, France}
\affiliation{Universit\'{e} Paris-Saclay, CEA, LMCE, 91680 Bruy\`{e}res-le-Ch\^{a}tel, France}
\author{L.~Gremillet}
\affiliation{CEA, DAM, DIF, 91297 Arpajon, France}
\affiliation{Universit\'{e} Paris-Saclay, CEA, LMCE, 91680 Bruy\`{e}res-le-Ch\^{a}tel, France}
\author{M.~Tamburini}
\affiliation{Max-Planck-Institut f\"ur Kernphysik, Saupfercheckweg 1, D-69117 Heidelberg, Germany}
\author{S.~Corde}
\email[Corresponding authors:\\ ]{aime.matheron@polytechnique.edu\\ pablo.san-miguel-claveria@polytechnique.edu\\ sebastien.corde@polytechnique.edu}
\affiliation{LOA, ENSTA Paris, CNRS, Ecole Polytechnique, Institut Polytechnique de Paris, 91762 Palaiseau, France}

\begin{abstract}
{\textbf{Abstract}}
\newline
Ongoing progress in laser and accelerator technology opens new possibilities in high-field science, notably to investigate the largely unexplored strong-field quantum electrodynamics (SFQED) regime where electron-positron pairs can be created directly from light-matter or even light-vacuum interactions. Laserless strategies such as beam-beam collisions have also been proposed to access the nonperturbative limit of SFQED. Here we report on a concept to probe SFQED by harnessing the interaction between a high-charge, ultrarelativistic electron beam and a solid conducting target. When impinging onto the target surface, the beam self fields are reflected, partly or fully, depending on the beam shape; in the rest frame of the beam electrons, these fields can exceed the Schwinger field, thus triggering SFQED effects such as quantum nonlinear inverse Compton scattering and nonlinear Breit-Wheeler electron-positron pair creation. Through reduced modeling and kinetic numerical simulations, we show that this single-beam setup can achieve interaction conditions similar to those envisioned in beam-beam collisions, but in a simpler and more controllable way owing to the automatic overlap of the beam and driving fields. This scheme thus eases the way to precision studies of SFQED and is also a promising milestone towards laserless studies of nonperturbative SFQED.
\end{abstract}

\maketitle

\noindent
{\textbf{Introduction}}
\newline
\noindent
Advances in multi-petawatt laser systems \cite{Danson_HPLSE_2015, Yoon_Optica_2021, Radier_HPLSE_2022} should soon allow strong-field quantum electrodynamics (SFQED) effects \cite{Nikishov_JETP_1964, Erber_RMP_1966, DiPiazza_RMP_2012} to be induced and probed in extreme-intensity laser-matter interactions \cite{Bell_PRL_2008, Ridgers_PRL_2012, Vranic_PRL_2014, Lobet_PRAB_2017, Magnusson_PRL_2019,  Vincenti_PRL_2019, Lecz_PPCF_2019, Blackburn_RMPP_2020, Mercuri_Baron_2021, Qu_PRL_2021, PhysRevLett.124.044801,BaumannQED}. The main experimental challenge will be to access, in a controlled manner, a regime characterized by a quantum parameter $\chi = E^*/E_{\rm cr}$ larger than unity, that is, where charged particles experience in their rest frame an electric field $E^*=\gamma \left \vert \mathbf{E_\parallel}/\gamma + \mathbf{E_\perp}+\mathbf{v} \times \mathbf{B} \right\vert  \simeq \gamma \left\vert \mathbf{E_\perp}+\mathbf{v} \times \mathbf{B} \right \vert$ stronger than the Schwinger critical field \cite{Schwinger_PR_1951} $E_\mathrm{cr} = m_e^2c^3/e\hbar \simeq 1.3\times 10^{18}\,\rm V\,m^{-1}$ ($m_e$ is the electron mass, $e$ the elementary charge, $c$ the speed of light, $\hbar$ the reduced Planck constant, $\mathbf{v}$ the particle velocity, $\gamma$ its Lorenz factor $\gg 1$, $\mathbf{E_\parallel}$ and $\mathbf{E_\perp}$ the electric field components respectively parallel and perpendicular to $\mathbf{v}$, and $\mathbf{B}$ the magnetic field). Strong-field QED processes such as quantum nonlinear inverse Compton scattering (NICS), equivalent to strong-field quantum synchrotron radiation, and nonlinear Breit-Wheeler (NBW) electron-positron pair creation become prominent when $\chi \gtrsim 1$ \cite{Kirk_PPCF_2009, DiPiazza_RMP_2012}, up to the point of profoundly modifying the dynamics of physical systems subject to such conditions \cite{Nerush_PRL_2011, Neitz_PRL_2013, Gonoskov_PRL_2014, Lobet_PRL_2015, Grismayer_PRE_2017}. In recent years, using laser pulses of $\sim 10^{21}\,\rm W\,cm^{-2}$ intensity, it has already become possible to explore some features of NICS in the marginally quantum regime \citep{Poder_PRX_2018, Cole_PRX_2018, Blackburn_RMPP_2020}.

Different schemes have been proposed to achieve, and diagnose\cite{zhang2023signatures}, well-controlled conditions under which the quantum parameter approaches or exceeds unity.
The most common strategy relies on the head-on collision of an intense laser pulse with a high-energy electron beam. In the rest frame of the beam, the laser field strength is boosted by a factor of $\sim 2\gamma$, possibly leading to high $\chi$ values when combining an ultrarelativistic beam and an ultraintense laser pulse. The first experimental observation of strong-field QED effects in such a configuration was made at SLAC \cite{Burke_PRL_1997, Bamber_PRD_1999}, using a $\sim 47\,\rm GeV$ electron beam and a laser pulse focused to a mildly relativistic intensity of $\sim 0.5\times 10^{18}\,\rm W \,cm^{-2}$. This seminal experiment attained $\chi \approx 0.3$ and yielded a total of about one hundred Breit-Wheeler positrons~\cite{Burke_PRL_1997}. Since then, all-optical laser-beam schemes involving laser-wakefield-accelerated electron beams \cite{Tajima_PRL_1979, Faure_Nature_2004, Geddes_Nature_2004, Mangles_Nature_2004} have been considered as well \cite{Vranic_PRL_2014, Lobet_PRAB_2017, Blackburn_PRA_2017}, the first experimental tests of this configuration achieving $\chi \approx 0.25$ \cite{Cole_PRX_2018, Poder_PRX_2018}. These scenarios, though, are characterized by oscillatory driving fields, with tens of femtosecond timescales, and probably even longer given common intensity contrast issues. Such features should contribute to blurring the strong-field QED observables, and so seem non-optimal towards precision studies.

Progress in focusing and compressing high-energy accelerator beams \cite{Yakimenko_PRAB_2019}, notably using wakefield-induced plasma lenses \cite{PChen, JBRosensweig, JJSU} and energy-chirped beams \cite{Emma_APLP_2021}, can also be leveraged to reach strong-field QED conditions\cite{ DelGaudio_PRAB_2019, Tamburini_PRD_2021}.
Thus, Yakimenko \textit{et al.} \cite{Yakimenko_PRL_2019} recently proposed to collide two ultrarelativistic ($\sim 100\,\rm GeV$), highly focused ($\sim 10\,\rm nm$), high-charge ($\sim 1\,\rm nC$) electron beams to achieve $\chi \gg 1$, and ultimately enter the nonperturbative SFQED regime characterized by $\alpha_{\rm f} \chi^{2/3}\gtrsim 1$ ($\alpha_{\rm f} = 1/137$ is the fine-structure constant) in which the conventional theoretical framework of strong-field QED breaks down \cite{Ritus_AP_1972, Narozhny_PRD_1980, Fedotov_JPhCS_2017, Podszus_PRD_2019, Ilderton_PRD_2019, Mironov_PRED_2020}. In this scenario, the self fields of one beam act for the other beam as the laser field in a laser-beam collision, yet with the significant advantage that those fields are half-cycle-like and sub-femtosecond, thus more adapted, in principle, to clean strong-field QED measurements. This scheme, however, necessitates extreme beam parameters and is highly sensitive to the very precise beam alignment required to ensure proper overlap and stable head-on collisions. In the likely event of shot-to-shot jittering, one should rely on the detected gamma-ray and positron spectra to infer, via a model-dependent deconvolution, the actual interaction conditions, a hurdle in the way of developing a reliable testbed of strong-field QED models.

\begin{figure}[t!]
    \centering
    \includegraphics[scale=0.32]{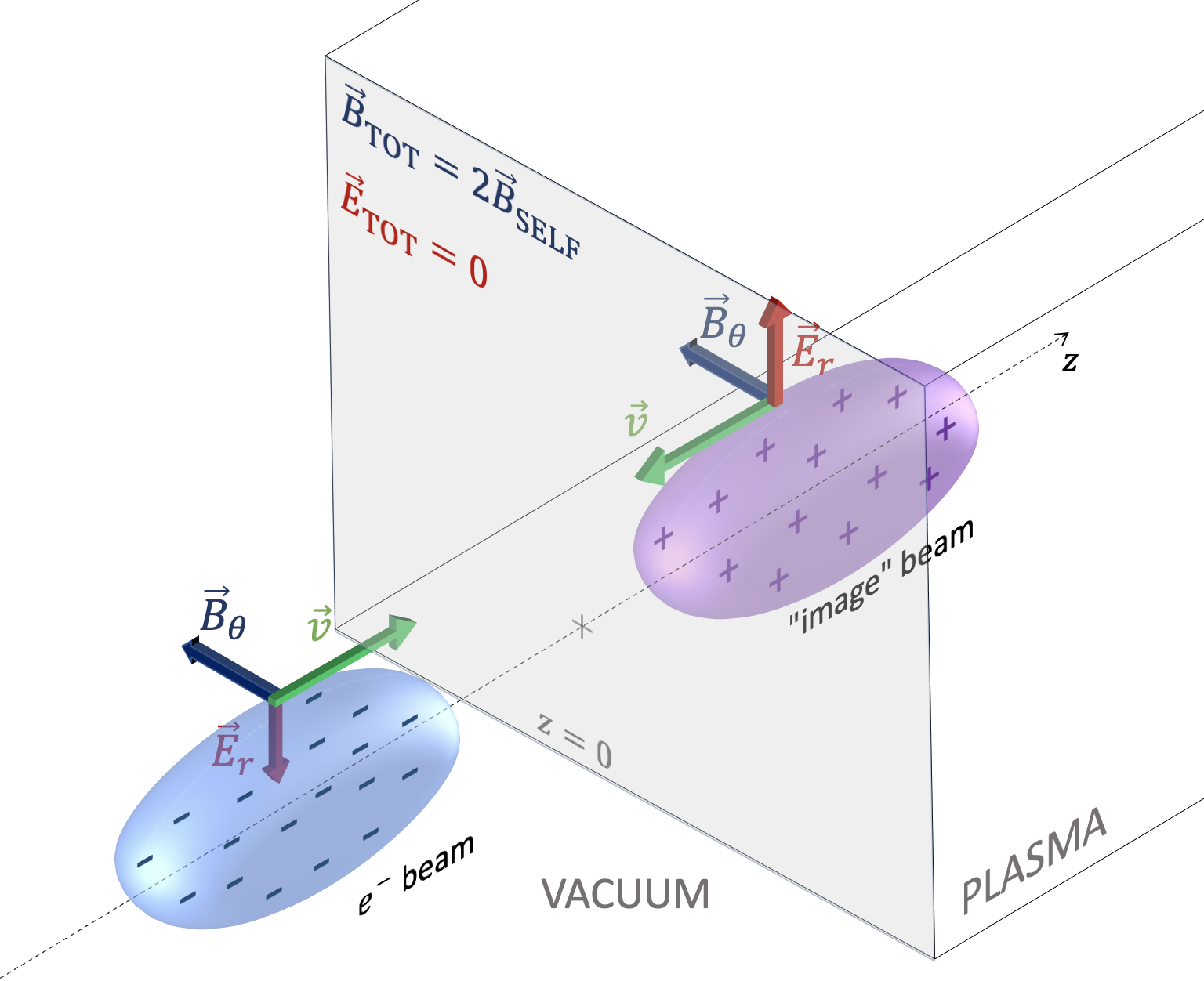}
    \caption{\textbf{Beam-plasma concept to probe strong-field QED.} An ultrarelativistic, high-density electron beam (in blue) entering a solid experiences the intense self fields of its image charge (in magenta) so that at z = 0, the electric field vanishes while the magnetic field doubles.}
    \label{fig1}
\end{figure}

Here, as a simpler alternative, we propose to employ a single ultrarelativistic, highly focused and compressed beam and make it interact with its own self fields that are reflected (in a broad sense) off the surface of a solid-density plasma \cite{Sampath_PRL_2021}. Under certain conditions, to be detailed later, this scenario can be viewed as a beam-beam collision\cite{DelGaudio_PRAB_2019, Tamburini_PRD_2021} in which the colliding beam is replaced with the incoming beam's image charge on the target (see Fig.~\ref{fig1}). In this simple picture, the total radial electric field vanishes at the plasma surface, whereas the azimuthal magnetic fields of the incoming and image beams add up. The microscopic sources of the image beam fields in the $z \leq 0$ space are the induced surface plasma currents that act to screen the incident beam self fields inside the solid-density plasma ($z>0$). Still, compared to beam-beam collisions, the beam-plasma scheme is much easier to implement as it does not involve a second, perfectly aligned electron beam. Even more importantly, it guarantees the overlap of the beam with the plasma-induced fields. Therefore, knowing the initial beam parameters, one can predict with high accuracy the time evolution of the fields, and hence of the $\chi$ parameter experienced by each beam particle, a key advantage for well-controlled strong-field QED measurements.

The objective of this work is to demonstrate, through a detailed analysis based on theoretical modeling and advanced particle-in-cell (PIC) simulations, the potential of ultrarelativistic beam-plasma collisions for precision strong-field QED studies. Besides the strong control over the interaction conditions inherent to this scheme, the requirement is to reach $\chi$ values well above unity, and over time scales short enough that each beam electron emits on average less than one gamma-ray photon. This ensures that the time history of the $\chi$ parameter is only determined by the initial beam parameters and not by the beam evolution during the interaction, which would complicate the benchmarking of strong-field QED models.

Our paper is structured as follows. First, we present a proof-of-concept PIC simulation that reproduces the results of the beam-beam setup. We then analyze the constraints posed to the beam shape and electron target density in order to achieve reflection of the beam self fields. Next, considering the case of an electron beam of fixed energy (10~GeV) and charge (1~nC) colliding with a gold foil target, we identify the optimal beam density and aspect ratio to reach high $\chi$ values. Finally, we demonstrate that the gamma-ray photon and positron spectra generated in this configuration can provide unambiguous experimental signatures of strong-field QED, that is, of the nonlinear inverse Compton scattering and nonlinear Breit-Wheeler processes that are found to largely prevail over the competing, undesirable Bremsstrahlung and Bethe-Heitler/Coulomb Trident processes.

\bigskip
\noindent
{\textbf{Results}}

\noindent
\textbf{Proof-of-concept PIC simulation.} 
\begin{figure*}[ht!]
    \centering
    \includegraphics[width=17cm]{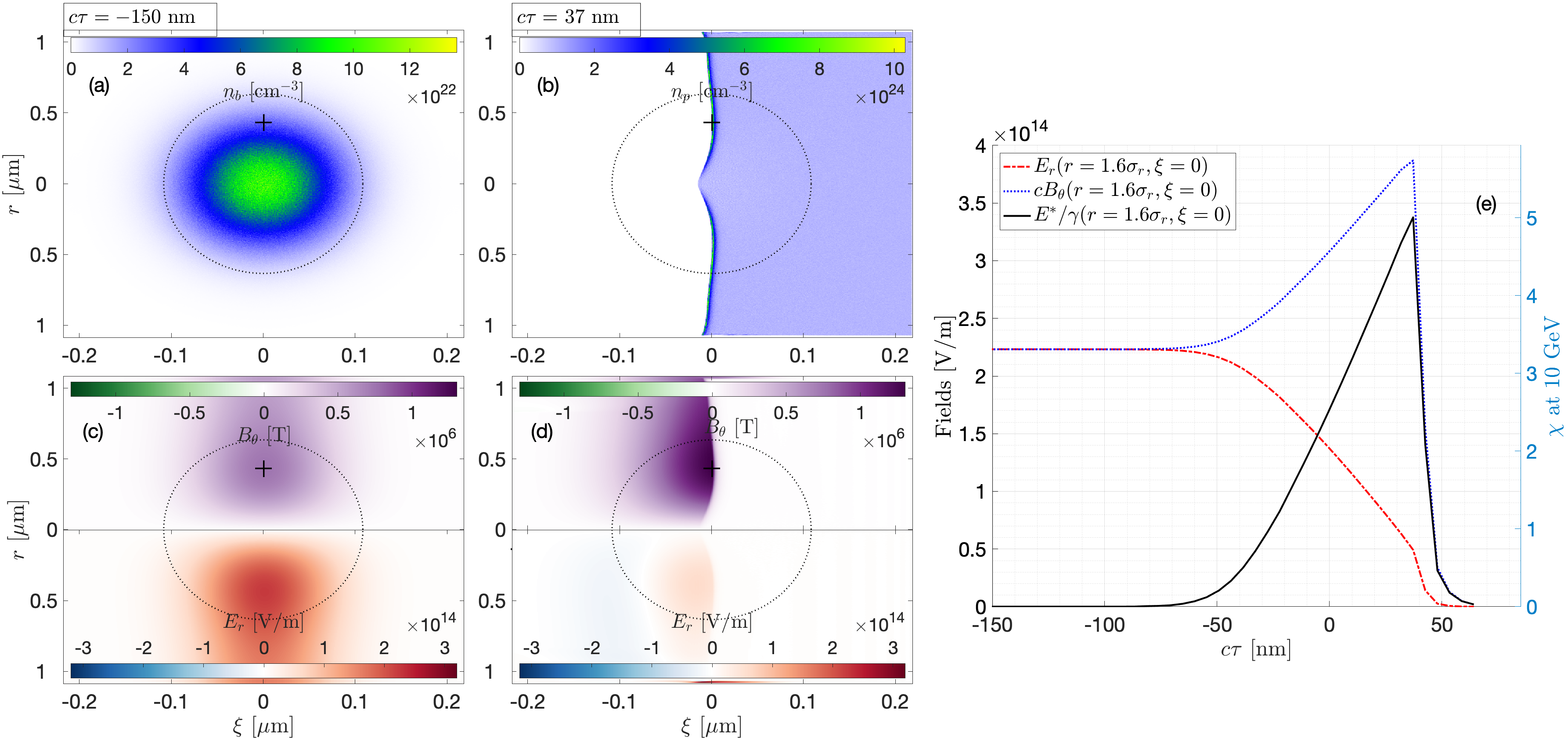}
    \caption{\textbf{Proof-of-concept 3D PIC simulation of a beam-plasma collision.} 2D ($\xi,r$) maps of beam density at $c\tau=-\SI{150}{nm}$ where the beam has not yet reached the plasma (a), of plasma electron density at $c\tau=\SI{37}{nm}$ where the beam has just entered the plasma (b), and of transverse electric and magnetic fields at $c\tau=-\SI{150}{nm}$ (c) and at $c\tau=\SI{37}{nm}$ (d). (e) Temporal evolution of the fields $E_r$, $cB_{\theta}$ and $E^*/\gamma$ at $(r, \xi)=(1.6\sigma_r,0)$. In (a)-(d), the black dotted circle represents the beam density contour at $5\%$ of maximum density and the black cross represents the position $(r, \xi)=(1.6\sigma_r,0)$ at which the fields are evaluated in (e).
    }
    \label{Fig2}
\end{figure*}
To assess the validity of the simple ``image beam'' picture, we have simulated the beam-plasma interaction using the three-dimensional (3D) PIC \textsc{calder} code \cite{Lefebvre_NF_2003}. Because the beam and the plasma have rotational symmetry about the propagation axis $z$, cylindrical coordinates will be used throughout the paper with ($r,\theta,z$) denoting the radial, azimuthal and longitudinal components respectively. Furthermore, since the simulation uses a moving window, the longitudinal comoving coordinates ($\xi=z-ct,\tau=t$) will be adopted. The beam center position is set at $\xi=0$, and $\tau=0$ corresponds to the time at which the central slice of the beam crosses the plasma surface. In this first simulation, the 10~GeV beam has a total charge of $Q=\SI{1}{nC}$ and a Gaussian density profile of rms length $\sigma_z=\SI{54}{nm}$ and width $\sigma_r=\SI{271}{nm}$. The peak beam density is then $n_{b,\rm peak} \simeq  10^{23}\,\rm cm^{-3}$. 
The beam is initialized together with its self fields in vacuum. The plasma target has a sharp boundary and a uniform electron density $n_p = 10^{24}\,\mathrm{cm}^{-3} \simeq 10 n_{b,\rm peak}$, a value typical of fully ionized solid foils. The simulation parameters are further detailed in the Methods section.

Figures~\ref{Fig2}(a) and \ref{Fig2}(b) show 2D ($\xi-r$) density slices of the beam in vacuum (a) and of plasma electrons when the beam has just entered the plasma (b). The corresponding maps of the radial electric field $E_r$ and azimuthal magnetic field $B_\theta$ are displayed in Figs.~\ref{Fig2}(c) and \ref{Fig2}(d). In Fig.~\ref{Fig2}(e) is plotted the temporal evolution of the fields as seen by a beam electron located at $(r,\xi)=(1.6\sigma_r, 0)$. This position, indicated by a cross in Figs.~\ref{Fig2}(a)-(d), is where the beam self fields are the highest. As expected, the beam self fields are fully screened inside the dense plasma: the $E_r$ field nearly vanishes at the surface while the $B_\theta$ field is approximately doubled. 

Figure~\ref{Fig2}(e) also shows the evolution of $E^*/\gamma \simeq E_r-cB_\theta$, which represents the net radial focusing force experienced by a beam electron. This quantity provides a figure of merit for strong-field QED effects that is independent of $\gamma$ and allows  $\chi=E^*/E_\mathrm{cr}$ to be directly evaluated when a particular $\gamma$ is specified. Note that the self fields of a relativistic beam (and thus the reflected fields) are independent of $\gamma$, so that $\chi$ simply scales as $\gamma$.

Our simulation confirms that, as in the beam-beam scenario, but in a much easier to realize and more robust setup, the beam-plasma configuration allows beam electrons to probe transient electromagnetic fields of amplitude similar to the beam self fields, which can trigger interactions with $\chi \gg 1$ at high Lorentz factors. Initially, when propagating in vacuum, the beam electrons are subjected to a negligible transverse force ($E^* \simeq 0$) as the transverse electric and magnetic self fields closely compensate each other in the laboratory frame. Just prior to entering the plasma, the beam electrons experience the reflected fields which impart a transverse force of magnitude comparable with that of the self fields. For a beam energy of $10\,\rm GeV$, this would translate into an electron quantum parameter $\chi \simeq 5$. However, as soon as the beam penetrates the plasma, $E^*/\gamma$ rapidly drops due to plasma shielding. Thus, akin to the beam-beam scheme, the beam electrons undergo a high $\chi$ value over a very brief duration only, of the order of $\sigma_z/c$, which is beneficial for precision studies of strong-field QED. 

\begin{figure*}[t!]
    \includegraphics[width=17cm]{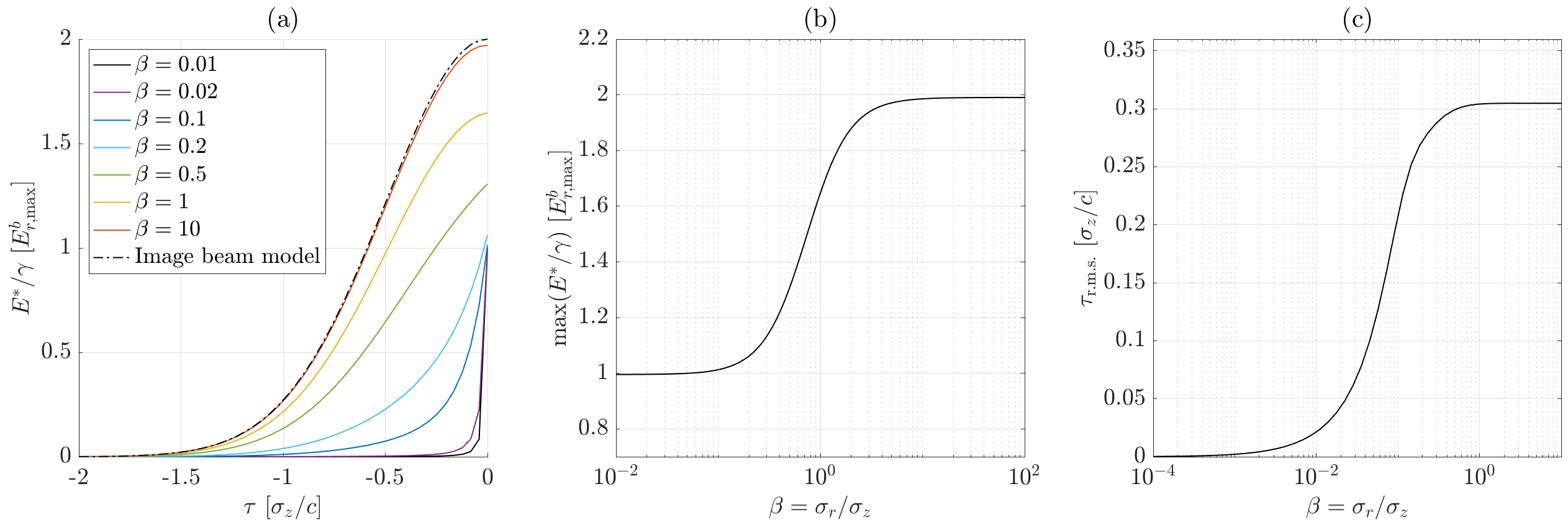}
    \caption{\textbf{Reflected and evanescent fields calculated with the pseudo-photon method.} (a): Temporal evolution of $E^*/\gamma$ in units of $E_{r,\mathrm{max}}^b$, as experienced by a beam electron located at $(r, \xi)=(1.6\sigma_r,0)$ just before it crosses the surface (corresponding to $c\tau=0)$. The dotted line represents the result of the ``image beam'' model, where the fields are assumed to be those of the image charge. (b): Maximum value of $E^*/\gamma$ in units of $E_{r,\mathrm{max}}^b$ and as a function of $\beta$. (c): Typical duration $\tau_\mathrm{rms}$ over which a beam electron experiences the field as a function of $\beta$.}
    \label{Fig3}
\end{figure*}

\bigskip
\noindent
\textbf{Propagating or evanescent fields.}
The properties of the reflection process are determined by the beam-to-plasma-density ratio $\alpha=n_{b,\mathrm{peak}}/{n_p}$ and the beam width-to-length ratio $\beta=\sigma_r/\sigma_z$. A necessary condition for the plasma to efficiently screen the self fields (yielding a nearly vanishing electric field at the surface) is that the electron plasma density is much larger than the beam density, that is, $\alpha  \ll 1$. Yet, even if the condition of vanishing electric field at the surface is satisfied, the image beam model only applies in the so-called radiating regime \cite{Sampath_PRL_2021} associated with a beam larger in width than in length, that is, with $\beta \gg 1$ (pancake-shaped beams). Then, the reflection process generates fields that propagate (in the direction of negative $z$) and diffract over an approximate Rayleigh length $\sigma_r^2/\sigma_z\gg\sigma_z$. Diffraction is negligible during the collision so that these propagating fields can be approximated to those of the image charge. In the nonradiating regime $\beta \ll 1$ (cigar-shaped beams),  the plasma-induced fields are evanescent, and so confined to the vicinity of the vacuum-plasma boundary. This configuration, not captured in the image beam approach, was already considered for focusing purposes several decades ago \cite{Adler_1982, Humphries_1988, Fernsler_1990}. A unified treatment of those two limits is provided by the pseudo-photon model developed by Sampath {\it et al.}\cite{Sampath_PRL_2021}, which we will exploit in the remainder of this Section.

In Fig.~\ref{Fig3}(a), the evolution of $E^*/\gamma$ as experienced by a beam electron located at $(r, \xi)=(1.6\sigma_r,0)$ and as computed by the pseudo-photon method, is represented for different size-to-length ratios $\beta$. While the agreement with the image beam model is excellent for $\beta=10$, one can see that both the maximum value of $E^*/\gamma$ and its duration (normalized to the beam length $\sigma_z$) are reduced at lower $\beta$, as a consequence of the induced fields turning evanescent. These two effects are quantified in Figs.~\ref{Fig3}(b) and \ref{Fig3}(c). First, in the nonradiating regime $\beta\ll1$, the maximum field amplitude experienced by a beam electron is equal to the maximum of the beam self electric field $E_{r,\rm max}^b$, in contrast with the doubled amplitude predicted by the image beam model, valid in the radiating regime $\beta \gg 1$. The transition from $\max(E^*/\gamma)=2E_{r,\rm max}^b$ to $\max(E^*/\gamma)=E_{r,\rm max}^b$ occurs around $\beta \simeq 1$ [see Fig.~\ref{Fig3}(b)], with $\beta=1$ (spherical beam) providing a reasonably high peak field strength, $\max(E^*/\gamma)=1.65E_{r,\rm max}^b$. Second, we observe that the duration $\tau_\mathrm{rms}$ over which a beam electron is subjected to a high value of $E^*/\gamma$ (see Methods for the definition of $\tau_\mathrm{rms}$) diminishes when the plasma-induced fields become increasingly evanescent. The $c\tau_{\rm rms}/\sigma_z$ ratio evolves from a constant value $\simeq 0.3$ at $\beta > 1$ [with $E^*/\gamma$ having a half-Gaussian shape in Fig.~\ref{Fig3}(a), corresponding to a propagating field] to smaller values at $\beta \ll 1$ [with $E^*/\gamma$ showing an exponential decay in Fig.~\ref{Fig3}(a), corresponding to an evanescent field], with $c\tau_{\rm rms}/\sigma_z \simeq 0.22$ at $\beta \simeq 0.1$.

The pseudo-photon model \cite{Sampath_PRL_2021} therefore predicts that the highest field $E^*/\gamma$ and highest duration $\tau_\mathrm{rms}$ are reached in the radiating regime, $\beta \gg 1$, but also that comparable performance can be obtained for $\beta\sim1$. However, using cigar-shaped beams with $\beta < 0.1$ should considerably reduce the time (in units of $\sigma_z/c$) over which the fields are experienced. 

\bigskip
\noindent
\textbf{Plasma screening and transparency.}
Given the scaling of the beam self field at fixed beam charge $Q$, $E^b_{r,\rm max} \propto \alpha^{2/3}\beta^{1/3}$, it seems desirable to increase $\alpha$ to boost the reflected field. However, when the beam density approaches the plasma density ($\alpha \sim 1$), the plasma is no longer able to perfectly screen the beam self field, which in turn may hinder the expected enhancement of the reflected field.

\begin{figure}[b!]
    \centering
    \includegraphics[width=8.5cm]{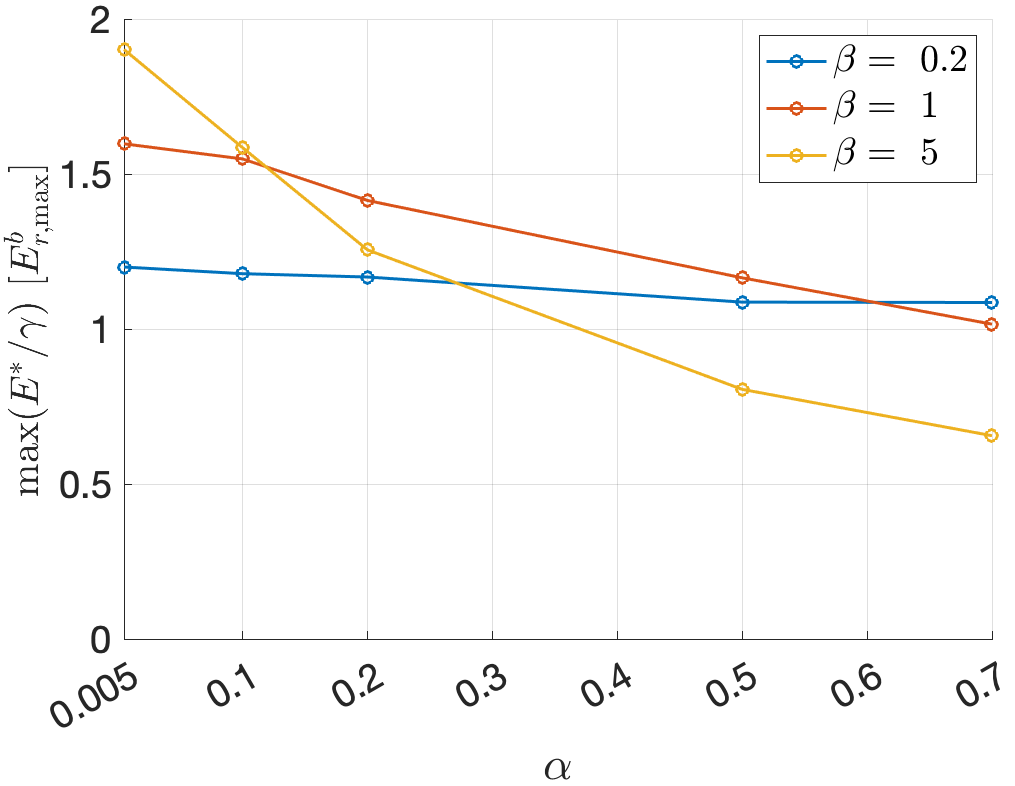}
\caption{\textbf{Impact of imperfect plasma screening.} Maximum value of $E^*/\gamma$ in units of $E_{r,\mathrm{max}}^b$ and evaluated at $(r, \xi)=(1.6\sigma_r,0)$ for different beam parameters $\alpha$ and $\beta$, for $Q=\SI{1}{nC}$ and $n_p=\SI{1e24}{cm^{-3}}$. Each data point represents the result of a PIC simulation. }
    \label{fig4}
\end{figure}

To evaluate the impact of imperfect plasma screening, we have carried out PIC simulations using the quasi-3D \textsc{calder-circ} code\cite{Lifschitz_JCP_2009} (see Methods), varying $\alpha$ from 0.005 to 0.7 for $\beta=0.2$, 1 and 5, and keeping the charge constant at $Q=\SI{1}{nC}$ and the plasma density at $n_p=\SI{1e24}{cm^{-3}}$. Figure~\ref{fig4} shows that, for $\alpha = 0.005$, the lowest simulated value, the PIC results reasonably agree with the pseudo-photon model [see Fig.~\ref{Fig3}(b)], with the maximum of $E^*/\gamma$ going from $1.9E_{r,\rm max}^b$ at $\beta=5$ (propagating fields for pancake-shaped beams) to $1.2E_{r,\rm max}^b$ at $\beta=0.2$ (evanescent fields for cigar-shaped beams).

\begin{figure*}[t!]
    \centering
    \includegraphics[width=17cm]{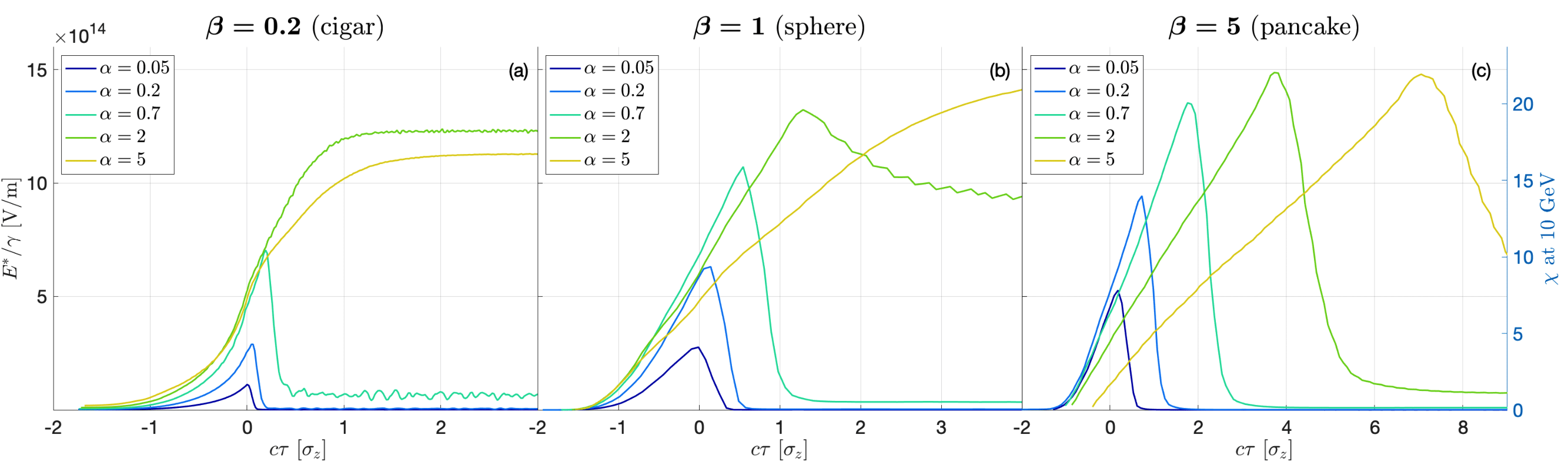}
    \caption{\textbf{Parametric study in a field-ionized gold target.} Temporal evolution of $E^*/\gamma$ (left scale) evaluated at $(r, \xi)=(1.6\sigma_r,0)$, and the corresponding $\chi$ value for a beam energy of $\SI{10}{GeV}$ (right scale), for $\beta=0.2$ (a), $\beta=1$ (b) and $\beta=5$ (c). Different values of $\alpha$ correspond to different beam densities $n_{b,\rm peak}$, and here the definition $\alpha=n_{b,\rm peak}/(60n_{p,1})$ is used for simplicity. Each curve corresponds to a separate PIC simulation.
    }
    \label{fig5}
\end{figure*}

However, when $\alpha$ is increased, the PIC results deviate substantially from the pseudo-photon model, which assumes perfect screening and is therefore independent of $\alpha$. At $\alpha \simeq 0.1$, the $\beta=1$ and $\beta = 5$ curves cross each other, and at $\alpha \simeq 0.7$, the hierarchy of $E^*/\gamma$ with $\beta$ is fully reversed. When going to $\alpha\gtrsim 1$, the reflection is strongly degraded and the interaction enters the blowout regime \cite{Rosenzweig_PRA_1991, Lu_PRL_2006} where the plasma electrons are fully expelled from the beam path, causing an ion cavity to form in the wake of the beam (see Supplementary Note 1). This regime, which can continue in the bulk of the plasma and is accounted for in our parametric simulation study below, starkly departs from the concept proposed here, based on the ultrashort surface interaction of the beam with its reflected (in a broad sense) self field. In comparison, the blowout mechanism generally occurs over longer timescales (set by the foil’s thickness), during which both the fields acting upon the beam and its resultant phase space may vary significantly, and not in a well-controlled way. Under our conditions, the nonlinear beam dynamics is made even more complex due to the abundant creation of electron-positron pairs that is enabled by the longer interaction time. These charged particles can strongly alter the fields experienced by the beam electrons, thus further hampering precision investigations of strong-field QED.

Figure~\ref{fig5} shows that working around $\alpha \simeq 0.5$ can provide a beam-plasma collision optimized for high $\chi$, with a reasonably high maximum value of $E^*/\gamma \simeq E^r_{r,\rm max}$ and a weak dependence on $\beta$. Beam-plasma collisions are therefore constrained by the achievable plasma density of the target and the need of $\alpha$ to stay below 1, which together limit $E^*/\gamma$.

The efficiency of the plasma shielding is not only determined by the beam-to-plasma density ratio but also to the response time of the plasma electrons $\sim \omega_p^{-1}$ ($\omega_p$ is the electron plasma frequency) compared to the beam duration $\sigma_z/c$. 
When $\omega_p\sigma_z/c \lesssim 1$,
i.e., when the beam duration is shorter than, or comparable with the inverse plasma frequency (or, equivalently, when the beam length is shorter than the plasma skin depth $k_p\equiv \omega_p/c)$, the plasma electrons have insufficient time to react to the beam self fields, which hence cannot be properly screened.
As the plasma turns transparent to the self fields, the maximum value of $E^*/\gamma$ is strongly reduced, with, for example, $\max(E^*/\gamma)\simeq0.1E_{r,\rm max}^b$ at $k_p\sigma_z=0.5$ (see Supplementary Note 2). This limit, however, is very hard to reach in practice given the extremely thin skin depth of solid-density plasmas ($k_p^{-1} \simeq 5\,\rm nm$ at $n_p=\SI{1e24}{cm^{-3}}$). Therefore, all results reported here use much longer bunches such that $k_p\sigma_z>1$. 

\bigskip
\noindent
\textbf{Target ionization and parametric study.}
To probe strong-field QED at high $\chi$ values, the beam-plasma scheme requires an electron beam with large self fields and thus of high density, yet lower than that of the plasma electrons to ensure efficient shielding and reflection of the beam self fields. The plasma should thus be as dense as possible. A conductor with free electrons can be a natural choice for the solid target. However, by leveraging the possible ionization of bound electrons in a high-$Z$ material by the beam self fields, higher electron plasma densities can be envisioned. 

To maximize the effective plasma electron density, we have considered a gold target ($Z=79$), whose atoms have one free electron (giving an electron plasma density $n_{p,1^+}=\SI{5.9e22}{cm^{-3}}$) and can provide, via field ionization, up to 76 additional electrons (giving a plasma density $n_{p,77^+}=\SI{4.5e24}{cm^{-3}}$) for the beam parameters considered here. We expect that the denser the beam, the stronger its self fields, and the higher the target ionization will be. A drawback of field ionization, however, is to make the relative beam density ($\alpha$) vary dynamically as a function of the instantaneous plasma electron density, which may increase the complexity of the problem.

To evaluate the induced fields and $\chi$ values that can be achieved in a gold target for different beam parameters, we have carried out a set of PIC simulations using the quasi-3D \textsc{calder-circ} code\cite{Lifschitz_JCP_2009} in which field ionization \cite{Nuter_PoP_2011} is enabled (see Methods for numerical details). This parametric study was performed for beam densities ranging from \SI{1.77e23}{cm^{-3}} to \SI{1.77e25}{cm^{-3}} and $\beta \in (0.2,1,5)$, keeping the charge constant at $Q=\SI{1}{nC}$. The target was initialized as a semi-infinite, neutral plasma composed of $\rm Au^{1+}$ ions and free electrons of uniform density $\SI{5.9e22}{cm^{-3}}$. With these parameters, our simulations show that the beam is able to ionize the target surface atoms to $\rm Au^{57+}$ for $n_b=\SI{1.77e23}{cm^{-3}}$ and $\beta=1$, and to $\rm Au^{77+}$ for $n_b\geq\SI{1.06e25}{cm^{-3}}$ and $\beta \in (0.2,1,5)$ before the central slice of the beam hits the surface (see Supplementary Note 3). 

For simplicity, Fig.~\ref{fig5} shows the results of this parametric study using the definition $\alpha=n_{b,\rm peak}/n_{p,60^+}=n_{b,\rm peak}/60n_{p,1^+}$, (i.e. with respect to the electron density of a $\rm Au^{60+}$ plasma), a reasonable approximation for gold ionization between $\rm Au^{57+}$ and $\rm Au^{77+}$. The time evolution of $E^*/\gamma$ (and hence $\chi$) is observed to be sharply peaked for $\alpha \le 0.7$, a behavior characteristic of a beam-plasma collision where the dominant fields are those reflected by the surface or the evanescent fields in its vicinity. Curves with late-time plateaus, particularly visible for $\beta=0.2$ and $\alpha\geq2$, illustrate the case where the physics is essentially dominated by the blowout mechanism, whereby plasma electrons are fully expelled from the beam path, around the surface as well as in the bulk of the plasma, allowing intense plasma-induced fields to be maintained at large times. For $\beta=5$, the peak values at $\alpha \geq 2$ are not substantially increased with respect to $\alpha=0.7$, thus highlighting the degraded reflection process. Furthermore, in these cases ($\alpha \geq 2,\beta=5$), $E^*/\gamma$ rapidly drops after its maximum despite the formation of an ion cavity. The reason for this behavior is that the highest fields in the cavity created by a pancake-shaped, short beam are located behind it, and so hardly act upon it. In addition, the peak of $E^*/\gamma$ is substantially shifted from $c\tau=0$, due to the strong forward push undergone by surface plasma electrons when the beam enters the target \cite{Xu_PRL_2021}. This accounts for the compressed plasma surface that is seen in Fig.~\ref{Fig2}(b).

From this parametric study, we conclude that in order to achieve ultrashort surface interaction of the beam with the plasma-induced fields, comparable in strength with the beam self fields, optimal performance is obtained for $\alpha\in[0.2,0.7]$ and $\beta\in[0.2,1]$.
The quantum parameter experienced by the beam particles then shows a sharply peaked temporal variation (of typical duration $\sim \sigma_z/c$) with maximum values $\chi \simeq 5-15$ for a beam energy of \SI{10}{GeV}. Pushing to higher $\alpha$ or $\beta$ (a very challenging experimental objective given the difficulty of compressing beams to such levels), only provides small improvement or leads to a transition to the more complex, nonlinear blowout regime, detrimental to well-controlled investigations of strong-field QED.

\begin{figure*}[t!]
    \centering
    \includegraphics[width=16cm]{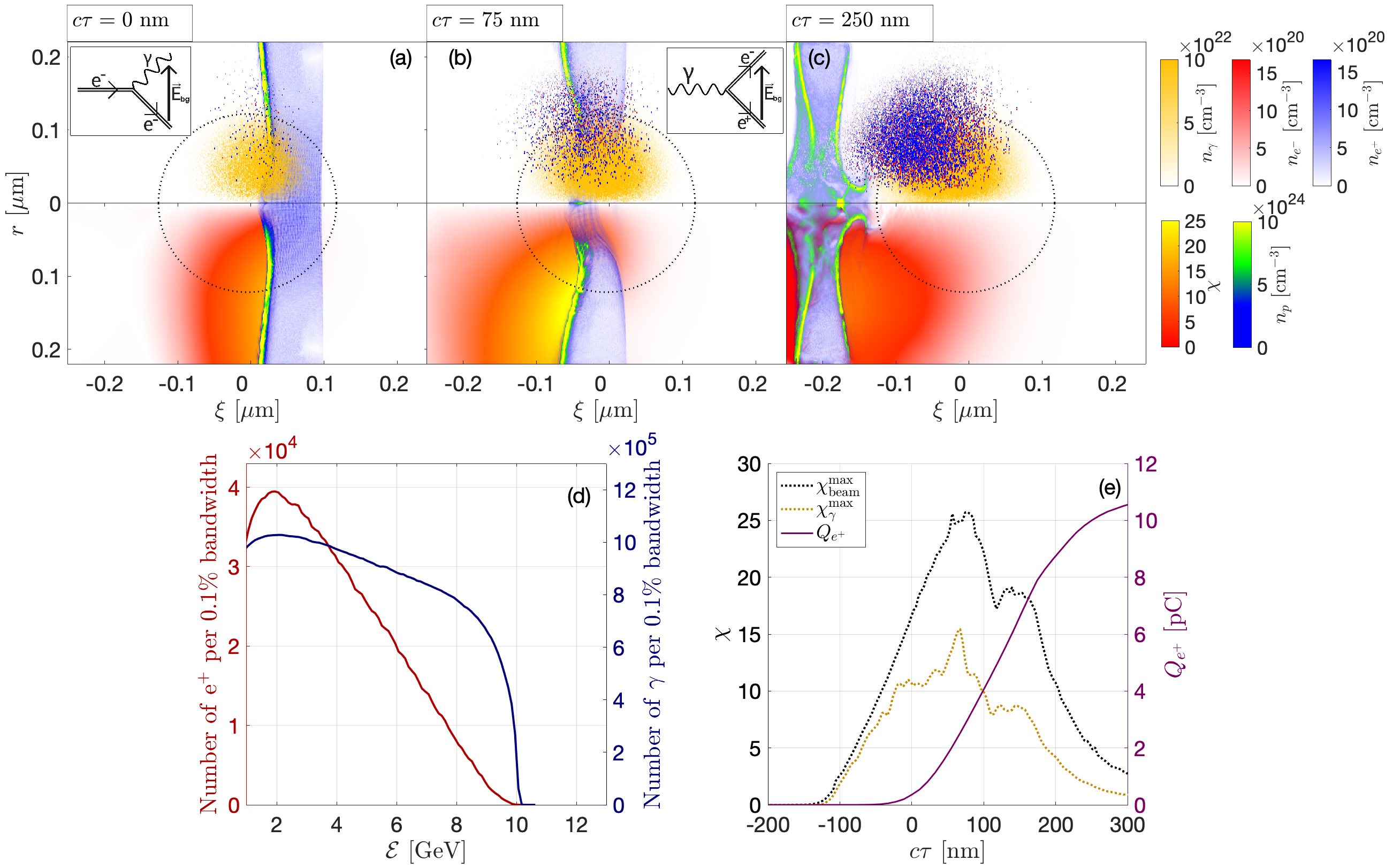}
    \caption{\textbf{Strong-field QED processes in the collision of a 10~GeV electron beam with a gold target.} Top: Spatial distributions of electron plasma density $(n_p)$, NICS $\gamma$-ray density $(n_\gamma)$, NBW electron $(n_{e^-})$ and positron $(n_{e^+})$ densities, and mean (beam electron) quantum parameter ($\chi$) at times $c\tau=0$ (a), $c\tau=\SI{75}{nm}$ (b) and $c\tau=\SI{250}{nm}$ (c). (d) $\gamma$-ray photon and positron spectra (photon and positron numbers per $0.1\,\%$ energy bandwidth) after the collision. (e) Temporal evolution of the maximum $\chi$ parameter experienced by beam electrons and $\gamma$ rays [maximum value taken inside the $5\%$ beam density contour shown as black dotted lines in (a)-(c)], and of the total charge of positrons above 1~GeV. The insets in (a) and (b) represent the Feynman diagrams for the NICS and NBW processes, respectively. Every sub figure uses the same colorbar and axes.}
    \label{fig6}
\end{figure*}

\bigskip
\noindent
\textbf{Strong-field QED observables.}
From the previous results, one can predict that the interaction of a 10~GeV electron beam characterized by $Q=\SI{1}{nC}$ and $\sigma_r=\sigma_z=\SI{55}{nm}$ ($\beta=1$) with a solid gold target ($\alpha=0.7$) will entail massive generation of $\gamma$-ray photons via NICS ($e^-+n\omega\rightarrow e^-+\gamma$) in the strong-field QED ($\chi \gg 1$) regime. In this process, a beam electron absorbs several low-energy photons ($n\omega$) from the electromagnetic field and emits a single $\gamma$ ray of high energy. The associated Feynman diagram is sketched in the inset of Fig.~\ref{fig6}(a), the double line representing the dressed state of the electron Dirac field in a strong electromagnetic field.

In turn, when subjected to the same driving fields as the beam electrons, the NICS $\gamma$ rays of energy $\varepsilon_\gamma > 2 m_e c^2$ can rapidly convert into electron-positron pairs via the NBW process ($\gamma+n\omega\rightarrow e^-+e^+$) if their quantum parameter, $\chi_\gamma = (\varepsilon_\gamma/m_e c^2) \vert \mathbf{E_\perp} + c\mathbf{\hat{k}_\gamma} \times \mathbf{B} \vert /E_{\rm cr}$, exceeds unity ($\mathbf{\hat{k}_\gamma}$ is the photon propagation direction) \cite{DiPiazza_RMP_2012}. The inset of Fig.~\ref{fig6}(b) shows the corresponding Feynman diagram.

To assess accurately the efficiency of these two strong-field QED processes, and confirm that they fully control the high-energy photon and positron emissions, we present in Fig.~\ref{fig6} the results of a 3D PIC-QED simulation of the above scenario in the case of a \SI{100}{nm}-thick gold foil. In addition to NICS and NBW \cite{Lobet_JPCS_2016}, the simulation describes the competing processes mediated by the Coulomb field of the target nuclei, namely, Bremsstrahlung together with Bethe-Heitler and Trident pair generation\cite{Martinez_PoP_2019}.

Upon crossing the foil surface, the beam electrons experience $\chi$ values as high as $\sim 25$ [see Figs.~\ref{fig6}(b) and~\ref{fig6}(e)], causing prolific NICS photon production, as expected. Figure~\ref{fig6}(a) shows that $\gamma$ rays are created with a doughnut-shaped distribution, which originates from the spatial distributions of the beam density and $\chi$ parameter. Figure~\ref{fig6}(d) further reveals that the $\gamma$-ray spectrum (photon number per $0.1\,\%$ energy bandwidth) is relatively flat and extends all the way up to the electron beam energy. It is to be emphasized that, despite the extreme fields at play, the interaction time $\sim \sigma_z/c$ is short enough that the average number of $\gamma$ rays emitted per beam electron remains lower than unity (this number is measured to be $\sim 0.46$ for photon energies $> 0.5\,\rm GeV$, i.e., $5\,\%$ of the energy of the individual beam electrons). The effective absence of successive high-energy photon emissions ensures that NICS occurs under well-controlled conditions (only determined by the initial beam parameters), and justifies {\it a posteriori} its neglect in our previous design study. Moreover, the total beam energy is found to drop only by $\sim 15\,\%$ during the interaction, essentially from NICS (the energy loss induced by the plasma-induced fields remains negligible). Such a weak variation in beam energy can further simplify the modeling of the problem.

Figure~\ref{fig6}(e) shows that the quantum parameter $\chi_\gamma$ of the NICS photons can reach values above $\sim 10$. Figures~\ref{fig6}(b) and~\ref{fig6}(c) detail the spatial distribution of the resultant NBW pairs during and after the passage of the electron beam through the target. The interaction results in the generation of a positron beam of $\gtrsim 10^{21}\,\rm cm^{-3}$ density and $\gtrsim 10\,\rm pC$ charge (above 1~GeV) that copropagates with the (approximately $10\times$ denser) $\gamma$-ray beam [see Figs.~\ref{fig6}(c) and \ref{fig6}(e)]. As for the photons, the energy spectrum of the positrons extends up to the initial beam energy, yet drops faster beyond its maximum at $\sim 2\,\rm GeV$ [Figs.~\ref{fig6}(d)].

Importantly, our advanced PIC-QED simulation further reveals that the competing (Bremsstrahlung, Bethe-Heitler and Coulomb Trident) processes that may generate undesirable $\gamma$-ray and positron background signals are all negligible for the thin (100~nm) gold target considered. In detail, the number of NICS $\gamma$ rays (NBW pairs) is about $10^3$ (resp. $1.5\times 10^3$) times higher than due to Bremsstrahlung (resp. Bethe-Heitler and Coulomb Trident processes). The outgoing $\gamma$-ray and positron spectra can therefore serve as unambiguous observables of the ultrashort-timescale, strong-field QED processes triggered during the beam-plasma collision, a result of prime significance for future experiments. 

\bigskip
\noindent
{\textbf{Conclusions}}

\noindent
Laboratory investigations of strong-field QED remain difficult nowadays due to the extreme electromagnetic fields and particle energies required to reach $\chi$ values well in excess of unity. We have demonstrated that the self fields of a very dense, ultrarelativistic charged particle beam can be harnessed, when coupled with an even denser plasma target, to achieve strong-field QED conditions. In particular, we have shown that a SLAC-class, 10~GeV, 1~nC electron beam with dimensions $\sigma_r=\sigma_z=\SI{55}{nm}$ can probe strong-field QED processes when colliding with a thin gold foil. The electron (photon) quantum parameter can reach values as high as $\chi \sim 25$ (resp. $\chi_\gamma \sim 15$), so that up to $\sim 10\,\rm pC$ of ultrarelativistic ($\ge 1\,\rm GeV$) positrons can be produced through NICS followed by NBW. The latter processes turn out to largely dominate other sources of $\gamma$ rays (Bremsstrahlung) and pairs (Bethe-Heitler, Coulomb Trident), and hence the beam-plasma scheme can provide unequivocal strong-field QED signatures, free of undesirable background.

As the $\chi$ parameter is proportional to the beam's Lorentz factor, increasing the latter would allow one to access even more extreme conditions (e.g. $\chi \sim 100$ at 40~GeV), or to relax the constraints on the beam parameters for a fixed $\chi$ value. For instance, given the $\alpha^{2/3} \beta^{1/3}$ scaling of the self field, the same $\chi$ value could be reached by going from 10~GeV to 40~GeV and lowering $\alpha$ by a factor of $(40/10)^{3/2}=8$, keeping $\beta$ and $Q$ constant.

While beam-beam collisions at future 100-GeV-class particle colliders could, in principle, enter the nonperturbative SFQED regime ($\chi \gtrsim \alpha_{\rm f}^{-3/2} \simeq 1600$) \cite{Yakimenko_PRL_2019}, their experimental realization appears highly challenging due to alignment issues. At this stage, the beam-plasma setup, which involves a single beam only and guarantees perfectly controlled interactions, seems to offer a much easier avenue to accelerator-based, fundamental strong-field QED studies in the $\chi\sim 10-100$ regime, and ultimately beyond by boosting the beam energy.

\bigskip
\noindent
{\textbf{Methods}}

{\small

\noindent
\textbf{Particle-in-cell simulations.} 
The 3D PIC code \textsc{calder} \cite{Lefebvre_NF_2003} (for Figs.~\ref{Fig2} and \ref{fig6}) and the quasi-3D PIC code \textsc{calder-circ} \cite{Lifschitz_JCP_2009} (for Figs.~\ref{fig4} and \ref{fig5}) were used to simulate the interaction between the electron beam and the plasma or gold target. In \textsc{calder-circ} runs, because of the axisymmetry of the problem, only the first mode ($m=0$) of the azimuthal decomposition was included, making it equivalent to a 2D $(r,z)$ simulation. The boundary conditions along the transverse direction were set to be reflective. The reason for this choice is that standard absorbing conditions were found unable to treat properly the beam self fields that propagate tangentially to the transverse boundary of the domain, and would extend, in actual conditions, well beyond it.
Reflective conditions proved satisfactory enough to reduce detrimental boundary errors.
All simulations reported in the paper used mobile ions but discarded Coulomb (elastic) collisions, which previous tests found of negligible influence.
The electron beam was initialized with zero temperature as emittance effects are negligible over the relatively short propagation distances considered. As the beam preserves its shape throughout the simulation, we only display it before entering the target in Fig.~\ref{Fig2}(a).
For the simulations shown in Figs.~\ref{Fig2}, \ref{fig4} and \ref{fig5}, in which the beam electrons propagate ballistically, each cell initially contained 2 plasma ions, 10 plasma electrons and 5 beam electrons. For the simulation depicted in Fig.~\ref{fig6}, each cell was filled with one plasma ion, 5 plasma electrons and 3 beam electrons, and the beam evolved self-consistently.

The 3D $(x,y,\xi\equiv z-ct)$ simulation domain associated with Fig.~\ref{Fig2} extends over $8\sigma_r \times 8\sigma_r \times 8\sigma_z$, and contains $1624 \times 1624 \times 650$ cells. The time step is of $\SI{1.8}{as}$ and the plasma has a uniform electron density $n_p=\SI{1e24}{cm^{-3}}$.
The 3D $(x,y,\xi)$ simulation domain corresponding to Fig.~\ref{fig6} has dimensions of $9\sigma_r \times 9\sigma_r\times 9\sigma_z$, discretized into $553 \times 553 \times 650$ cells. The time step is of $\SI{1.2}{as}$ and the plasma is initialized with Au$^{1+}$ ions and a uniform density of $\SI{5.9e22}{cm^{-3}}$. Synchrotron and Bremsstrahlung radiation, pair creation through nonlinear Breit-Wheeler, Bethe-Heitler and Coulomb Trident processes, and field ionization are described simultaneously \cite{Lobet_JPCS_2016, Martinez_PoP_2019}.

The quasi-3D simulations of Figs.~\ref{fig4} and \ref{fig5} consider a plasma initialized with Au$^{1+}$ ions and a uniform density of $\SI{5.9e22}{cm^{-3}}$. Field ionization is activated, each ionization event leading to a new electron macroparticle. Therefore, ionization from Au$^{1+}$ to Au$^{77+}$ produces 76 new electrons per gold ion. These simulations cover a $(r,\xi)$ domain of size $8\sigma_r \times 10\sigma_z$ with:
\begin{itemize}
    \item for $\beta=1$, a mesh made of $200 \times 500$ cells and time steps of $\SI{7.7}{as}$, $\SI{4.7}{as}$, $\SI{3.1}{as}$, $\SI{2.2}{as}$, $\SI{1.6}{as}$ for, respectively, $\alpha=0.05,\ 0.2,\ 0.7,\ 2,\ 5$.
    \item for $\beta=0.2$, a mesh made of $200 \times 2500$ cells and time steps of $\SI{4.3}{as}$, $\SI{2.8}{as}$, $\SI{1.8}{as}$, $\SI{1.3}{as}$, $\SI{0.9}{as}$ for, respectively, $\alpha=0.05,\ 0.2,\ 0.7,\ 2,\ 5$.
    \item for $\beta=5$, a mesh made of $1000 \times 500$ cells and time steps of $\SI{2.6}{as}$, $\SI{1.6}{as}$, $\SI{1.1}{as}$, $\SI{0.7}{as}$, $\SI{0.5}{as}$ for, respectively, $\alpha=0.05,\ 0.2,\ 0.7,\ 2,\ 5$.
\end{itemize}
\noindent
\textbf{Definition of the interaction time plotted in Fig.~\ref{Fig3}(c).}
The typical time $\tau_{\rm rms}$ over which a beam electron experiences the fields is defined as the rms duration:
\begin{equation*}
\tau_{\rm rms}=\sqrt{\frac{\int dt\ t^2 E^*/\gamma }{\int dt\ E^*/\gamma}-\left( \frac{\int dt\ t E^*/\gamma}{\int dt\ E^*/\gamma}\right)^2} \,.
\end{equation*}
}

\bigskip
\noindent
\textbf{Data availability}

\noindent
The data that support the findings of this study are available from the
corresponding author upon request.

\bigskip
\noindent
\textbf{Code availability}

\noindent
The CALDER code has been used to generate the results presented in this manuscript. This code has already been characterized with a full description reported in the references provided in the manuscript\cite{Lefebvre_NF_2003,Lifschitz_JCP_2009,Nuter_PoP_2011,Lobet_JPCS_2016,Martinez_PoP_2019}.

\bigskip
\noindent
{\textbf{References}}

\bigskip
\noindent
{\textbf{Acknowledgments}} 
This work was performed in the framework of the E-332 Collaboration. E-332 is a SLAC experiment which aims at the study of near-field coherent transition radiation in beam-multifoil collisions and the resulting self-focusing, gamma-ray emission and strong-field QED interactions. The work at CEA and LOA was supported by the ANR (UnRIP project, Grant No. ANR-20-CE30-0030). The work at LOA was also supported by the European Research Council (ERC) under the European Union's Horizon 2020 research and innovation programme (M-PAC project, Grant Agreement No. 715807). P. San Miguel Claveria was supported by the ANR under the programme “Investissements d’Avenir” (Grant No. ANR-18-EURE-0014). We acknowledge GENCI-TGCC for granting us access to the supercomputer IRENE under Grants No. 2020-A0080510786, 2021-A0100510786, 2021-A0110510062 and 2022-A0120510786 to run PIC simulations. J.~R.~Peterson was supported by DOE NNSA LRGF fellowship under grant DE-NA0003960. The work at SLAC was supported by U.S. DOE FES Grant No. FWP100331 and DOE Contract DE-AC02-76SF00515. UCLA was supported by U.S. Department of Energy Grant No. DE-SC001006 and NSF Grant No. 1734315.

\bigskip
\noindent
{\textbf{Author contributions}} 

\noindent
Simulations, analytical work and data analysis were carried out by A.M. and P.S., with the support and supervision of X.D., L.G., M.T. and S.C. A.M., P.S., R.A., H.E., F.F., S.G., M.G., M.H., C.K., A.K., M.L., Y.M., S.M., Z.N., B.O., J.P., D.S., Y.W., X.X., V.Z., X.D., L.G., M.T., S.C. discussed the results and their interpretation. A.M., P.S., L.G. and S.C. wrote the manuscript with inputs from all authors. S.C. supervised the project.

\bigskip
\noindent
{\textbf{Competing interests}} 

\noindent
The authors declare no competing interests.

\end{document}